\documentclass[letterpaper, 10 pt, conference]{ieeeconf}  % Comment this line out
\IEEEoverridecommandlockouts                              % This command is only
\usepackage{amsmath, amssymb}
\usepackage{comment}
\usepackage{graphicx}
\usepackage{url}
\usepackage{subcaption}
\usepackage{epstopdf}
\usepackage{xcolor}
\usepackage{booktabs}
\usepackage{tikz}
\usepackage{mathtools}

\allowdisplaybreaks % equations can break across pages

\usepackage{amsthm}
\newtheorem{remarkEnv}{Remark}

\newtheorem{lemmaEnv}{Lemma}
\newenvironment{lemma}[1][]{\begin{lemmaEnv}}{\hfill\end{lemmaEnv}}

\title{\LARGE \bf
Detectability of Subtle Anomalies in Dynamical Systems\\via Log-Likelihood Ratio
}
\author{Alejandro Penacho Riveiros, Matthieu Barreau, and Nicola Bastianello% <-this % stops a space
\thanks{This work was supported by the Wallenberg AI, Autonomous Systems and Software Program (WASP) funded by the Knut and Alice Wallenberg Foundation.}% <-this % stops a space
\thanks{The authors are with the Department of Decision and Control Systems, and Digital Futures, KTH Royal Institute of Technology, Stockholm, Sweden {\tt\small \{alejpr,barreau,nicolba\}@kth.se}.}%
}

\begin{document}

\maketitle
\thispagestyle{empty}
\pagestyle{empty}

%%%%%%%%%%%%%%%%%%%%%%%%%%%%%%%%%%%%%%%%%%%%%%%%%%%%%%%%%%%%%%%%%%%%%%%%%%%%%%%%
\begin{abstract}
Industrial control applications require detecting system anomalies as accurately and quickly as possible to enable prompt maintenance.
In this context, it is common to consider several possible plant models, each linked to a different anomaly. The log-likelihood ratio method can then be used to identify the most accurate model and thereby classify which anomaly, if any, has occurred.
%
%However, despite its history, this method lacks a clear application to real-time anomaly detection for dynamical systems.
%
Although the method has been applied to a wide variety of systems, there is no formal analysis of what makes anomalies more or less prone to detection.
In this paper, we investigate a real-time anomaly detector based on the log-likelihood ratio and provide a theoretical characterization of its error rate when it is applied to linear Gaussian systems.
We showcase the performance of this algorithm and the characterization obtained, and demonstrate how the latter can be leveraged for observer design.
\end{abstract}

% Many industrial applications require detecting changes in system properties as soon as possible. In such situations, it is common to consider several possible models of the plant and evaluate which one is closer to the actual system using the log-likelihood. Although this technique has proven to be effective, there is no analysis of what makes certain system variations more or less prone to being detected with this method. In this paper, we analyze the detectability of small deviations using the log-likelihood ratio and derive analytical expressions that characterize the number of observations required for reliable model discrimination.

%%%%%%%%%%%%%%%%%%%%%%%%%%%%%%%%%%%%%%%%%%%%%%%%%%%%%%%%%%%%%%%%%%%%%%%%%%%%%%%%
\section{Introduction}
% \begin{itemize}
%     \item Detecting deviations of a system from its nominal model can usually hint towards internal issues that need to be addressed.
%     \item Detecting such deviation can be done from the measurements obtained from, through sensors.
%     \item A common approach is to design a bank of observers tuned for different possible variations of a plant, and observe which one is more accurate, that is, has less deviation.
%     \item Another approach is to 
%     \item Although these methods have proven to be effective for most applications, their effectiveness is usually analyzed numerically.
%     \item In this paper, we wonder what makes a system deviation more or less detectable, that is, it leads to faster detection. 
%     \item The analysis can be used to analyze what deviations are easier to detect, which ones are easy to confuse, and how the output of a system can be processed to improve detection.
% \end{itemize}

Industrial process monitoring is essential for ensuring the safe and efficient operation of complex systems such as chemical plants, energy facilities, and manufacturing processes \cite{li_review_2025}. Advances in sensing technologies and the widespread adoption of Internet of Things (IoT) infrastructures have enabled the collection of large volumes of data from distributed components, providing new opportunities for real-time monitoring and decision-making \cite{reis_industrial_2017}.

The main objective of process monitoring is to detect deviations of the plant from nominal system behavior. Such deviations may arise from component degradation, environmental changes, or external disturbances \cite{noauthor_validating_2018}. These deviations can indicate the onset of critical faults, such as early-stage material degradation or sensor drift. Detecting these subtle anomalies, also called incipient faults, reliably and in a timely manner, is therefore a key requirement in modern monitoring systems \cite{ding_model-based_2013, dion_anticipation_2024, yang_incipient_2022}.

The problem of detecting deviations from nominal behavior has been addressed using numerous approaches of varying complexity. Early methods relied on statistical tests applied directly to measured data, often neglecting the underlying system dynamics \cite{grant_statistical_1996} and lacking real-time applicability. Subsequent developments have incorporated knowledge of the plant's dynamical model, leading to model-based fault detection techniques based on observers and residual generation \cite{frank_fault_1990,isermann_fault-diagnosis_2011}. More recently, data-driven approaches, such as machine learning and deep learning methods, have been proposed to capture complex patterns in large datasets without explicit model assumptions \cite{iqbal_fault_2019, rodriguez-ramos_integrated_2024}, often outperforming traditional approaches at the expense of explainability.

Despite the practical success of all these approaches, their performance is usually evaluated empirically, and limited attention has been given to quantifying the number of observations that allow to reliably detect different types of anomalies \cite{shalev-shwartz_understanding_2014}. This question becomes particularly important when deviations from nominal behavior are small, as in models with very similar statistical properties \cite{basseville_detection_1993}. In such cases, distinguishing between hypotheses is inherently difficult and may require a large number of measurements, which can be problematic in applications where rapid detection is needed, as is usually the case in changepoint detection problems \cite{gustafsson_adaptive_2000}.

In this paper, we study anomaly detection and classification when the models differ only slightly. Focusing on linear systems driven by Gaussian noise, we analyze how the magnitude and structure of deviations from nominal behavior affect their detectability, and how many observations are required to identify the correct model with a prescribed level of confidence. To this end, we employ the cumulative log-likelihood ratio \cite{wald_sequential_1945, kang_sequential_2023}, adapted to dynamical systems.
%
% which is still widely used today due to its theoretical properties \cite{cover_elements_2006}. To adapt it to dynamical systems, we exploit steady-state properties to obtain tractable analytical results.
%
Once we have adapted the cumulative log-likelihood ratio method to steady-state linear-Gaussian systems to detect and classify anomalies, our main contributions  are:
\begin{enumerate}
    %\item We adapt the cumulative log-likelihood ratio method to steady-state linear-Gaussian systems to detect and classify anomalies in real time.
    
    % derive a real-time anomaly detection and classification algorithm based on the cumulative log-likelihood ratio applied to the system's output.

    \item We characterize the statistical properties of the ratio, which allow us to derive an analytical bound to the number of output measurements required to detect and classify two competing anomalies.
    
    \item We provide a heuristic to approximate this analytical bound in practice, based on \textit{e.g.} historical data.

    \item We demonstrate the numerical detection performance of the algorithm. Additionally, we show how to leverage it for observer design, highlighting the important trade-off between observer accuracy and detectability.
\end{enumerate}

% \begin{enumerate}
%     \item We derive an analytical method to estimate the number of measurements required to discriminate between two competing models under a prescribed error probability.
%     \item We provide a quadratic approximation of the log-likelihood ratio that enables practical comparisons across different model deviations, and characterizes their relative detectability. We build a link between this approximation and Fisher information.
%     \item We demonstrate how these results can be leveraged to design observers that enhance the detectability of specific deviations. We highlight the trade-off that arises in this context, between accuracy of the observer and detectability.
% \end{enumerate}

The remainder of the paper is organized as follows. Section \ref{sec:problem} introduces the problem formulation and the cumulative likelihood-ratio. Section \ref{sec:static} applies likelihood-based detection to sequences of observations, deriving detectability properties as a function of deviations. This sets the stage for Section \ref{sec:dynamic}, which extends these results to dynamical systems. Section \ref{sec:experiments} applies the proposed method for detection and observer design, presenting numerical results. Conclusions and future work are discussed in Section \ref{sec:conclusions}.

%%%%%%%%%%%%%%%%%%%%%%%%%%%%%%%%%%%%%%%%%%%%%%%%%%%%%%%%%%%%%%%%%%%%%%%%%%%%%%%%
\section{Problem Statement and Preliminaries} \label{sec:problem}

We proceed by introducing the problem setup: first, defining the system whose deviations we want to detect; second, how these deviations are defined; and finally, how the cumulative log-likelihood ratio is used to detect them.

\subsection{Model description} \label{sec:model-description}
We consider a linear, discrete dynamical system $\Pi = (A, C, Q, R)$, with state $x \in \mathbb{R}^n$ and measurements $y \in \mathbb{R}^p$. Their evolution is given by
\begin{equation}\label{eq:system_model}
\begin{split}
  x_{k+1} &= A x_k + w_k \qquad w_k \sim \mathcal{N}(0,Q),
  \\
  y_k &= C x_k + v_k \qquad v_k \sim \mathcal{N}(0,R),
\end{split}
\end{equation}
where $A \in \mathbb{R}^{n \times n}$ is the transition matrix, $C \in \mathbb{R}^{p \times n}$ is the observation matrix, $Q \in \mathbb{R}^{n \times n}$ is the covariance of the process noise and $R \in \mathbb{R}^{p \times p}$ is the covariance of the observation noise. Both the process and observation noise are i.i.d. and independent from each other.

An important aspect of this paper is our focus on autonomous systems driven by zero-mean Gaussian noise, motivated by closed-loop systems that are stabilized around a fixed operating condition via state feedback.
A consequence of~\eqref{eq:system_model} is that $\mathbb{E}[x_k] = \mathbb{E}[y_k] = 0$ once the effect of initial conditions has dissipated. This rules out most anomaly detection techniques, designed to detect additive faults, and motivates the need for detection methods that pick up more subtle clues of deviation \cite{ding_model-based_2013}.

% An important property of the system is the lack of input, so it is purely driven by Gaussian noise with zero mean. For a closed loop system, this property is common when it is designed to stay at a fixed operating point, stabilized by feedback control. It then follows that $\mathbb{E}[x_k] = \mathbb{E}[y_k] = 0$ once the effect of initial conditions has dissipated. This is a fundamental element of the paper, since anomaly detection techniques typically relies on deviations in the trajectory of the system to detect changes in its behavior. 

An additional complication in our setup is that the actual system matrices are unknown. Instead, we have knowledge of a nominal plant model, defined as $\Pi_* = (A_*, C_*, Q_*, R_*)$, which is expected to be close to the actual plant in standard conditions. This nominal plant can be obtained through first-principles modeling, system identification, or a combination of both.
Our analysis then is based on characterizing the actual system's model via a set of possible deviations from the nominal model, each one \textit{e.g.} due to degradation of a different component.
Specifically, let the model deviation $i$ be characterized by $\Pi_i^\Delta = (A_i^\Delta, C_i^\Delta, Q_i^\Delta, R_i^\Delta)$, then the system is modeled as a linear combination of deviations w.r.t. $\Pi_*$:
\begin{align}
  \Pi_\gamma = &\Pi_* + \sum_{i=1}^{N_\Delta} \gamma_i \Pi_i^\Delta
  \triangleq 
    \left( A_* + \sum_{i=1}^{N_\Delta} \gamma_i A_i^\Delta,\right. \label{eq:pi_gamma}
    \\ &C_* + \sum_{i=1}^{N_\Delta} \gamma_i C_i^\Delta,
    \left. Q_* + \sum_{i=1}^{N_\Delta} \gamma_i Q_i^\Delta,
    R_* + \sum_{i=1}^{N_\Delta} \gamma_i R_i^\Delta
  \right), \nonumber
\end{align}
with weights $\gamma = [\gamma_1, \gamma_2, \dots, \gamma_{N_\Delta} ] \in \mathbb{R}^{N_\Delta}$.
As an example, consider a system whose $A$ depends on a parameter $\kappa$ representing the damping of some internal mechanism; the deviation related to a change in this parameter then is described as $\Pi_i = \left( (\partial A)/(\partial \kappa), 0, 0, 0 \right)$. 
We remark that no loss of generality is incurred by using this representation, since an arbitrary number of deviations can be accounted for.
In practice, the weights $\gamma$ are of course unknown, since we do not have access to the actual system's model. Rather, our detection approach will rely on testing different hypotheses about the values of these weights, selecting those that yield more accurate predictions of the system's behavior.

The detection problem we address in this paper is further complicated by the following features of the systems that we consider.
First of all, we assume that any deviated system $\Pi_\gamma$ remains stable, for example, because the system has a large stability margin. Therefore, detection cannot rely on the coarse differences in the system's behavior that would be caused by an unstable system diverging. We remark that a large stability margin is typically observed in industrial processes operating in safe regimes and using robust controllers.
Secondly, we assume that model deviations are small, that is, $\|\gamma\| \ll 1$.
These deviations are harder to detect as they have a less obvious impact on the system's behavior, thus requiring a specialized detection technique.
As we will show later, this detection technique requires a larger number of measurements to compensate for the small magnitudes of the deviations at play. Thus, characterizing a sufficient number of measurements for detectability is important.

% This assumption has two important consequences for the problem. The first is that since deviations are small, we can approximate many quantities in the problem by linearization. This simplifies the math and allows us to obtain clean and intuitive expressions for the detectability of deviations. The second consequence is that, since small deviations require a large number of measurements to be detected, we can focus our analysis on the regime of large numbers of measurements, that is, $N \rightarrow \infty$, which will be key for estimating the probability distribution of the cumulative log-likelihood ratio.

\subsection{Model discrimination using cumulative log-likelihood}
As discussed above, the actual deviation $\gamma$ is unknown.
Therefore, in this paper, we consider two hypotheses parameterized by deviations' weights $\alpha, \beta \in \mathbb{R}^{N_\Delta}$, and aim to determine which one better explains the sequence of measurements $\{y_k\}_{k=1}^N$ we collect from the system.
The metric used for this task is the cumulative log-likelihood, defined as
\begin{align*}
    \mathcal{L}(\{y_k\}_{k=1}^N |\alpha,\beta)
    =
    \sum_{k=1}^N \mathcal{L}(y_k |\alpha,\beta) \nonumber
    =
    \sum_{k=1}^N \log \frac{p(y_k|\alpha)}{p(y_k|\beta)},
\end{align*}
where $p(y|\alpha)$ is the probability of observing $\alpha$ given the model $\alpha$.
A positive log-likelihood ratio implies that $\alpha$ explains the data better than $\beta$, and therefore yields a more accurate representation $\Pi_\alpha$ of the actual system's model $\Pi_\gamma$.
In the remainder of the paper, $\beta$ will represent a model close to nominal, and $\alpha$ a model of anomalous behavior.

% In general, we use $\alpha$ for the nominal model and set it to $0$, while $\beta$ represents some expected deviation.

\subsection{Problem formulation}
We are now ready to formalize the problem of interest.
Consider a system represented by the unknown model $\Pi_\gamma$ in~\eqref{eq:pi_gamma}, based on a set of $N_\Delta$ possible deviations from the nominal model $\Pi_*$.
Let $\alpha, \beta \in \mathbb{R}^{N_\Delta}$ be hypotheses for the values of the unknown $\gamma$, with $\alpha$ representing anomalous behavior requiring corrective action.
The objective is to determine, via the cumulative log-likelihood, which of $\Pi_\alpha$ and $\Pi_\beta$ better predicts the system's model, alerting to the need for maintenance when it is the latter.
In particular, we want to characterize the probability that the log-likelihood score is positive or negative as a function of the number of measurements.

% In this paper, we want to analyze how the cumulative log-likelihood evolves depending on the properties of the competing models $\alpha$ and $\beta$, and the actual model of the system $\gamma$. In particular, we want to determine the probability of the score being positive or negative as a function of the number of measurements taken.

\smallskip

The proposed algorithm to solve this problem is presented in two steps. In section~\ref{sec:static} (\textsc{Step 1}), we start by showing how to apply the log-likelihood score to distinguish which distribution generated measurements $\{ y_k \}_{k = 1}^N$.
Section~\ref{sec:dynamic} (\textsc{Step 2}) will then extend this result to distinguish between dynamical models generating $\{ y_k \}_{k = 1}^N$.

% To analyze the algorithm, we will proceed in several steps. In \textsc{Step 1}, we will consider a static setting, in which the measurements $y_k$ are produced by a fixed covariance matrix $\Sigma_\gamma$, and we have two competing covariance matrices $\Sigma_\alpha$ and $\Sigma_\beta$. 
% To extend to the dynamic setting, we will compute in \textsc{Step 2} how small changes in the matrices $A$, $C$, $Q$, and $R$ are carried into the covariance of the measurements in steady state $\Sigma_\gamma^y$. Then, we will return to the theory developed for the static case to analyze the detectability of different changes in the model.

\section{Step 1: Log-likelihood Discrimination for Static Normal Distributions} \label{sec:static}
This section prepares the tools and results needed to design the proposed detection algorithm for dynamical systems.

\subsection{Log-likelihood discrimination of distributions}
We focus here on an i.i.d. sequence of measurements generated by the unknown distribution $y_k \sim \mathcal{N}(0, \Sigma_\gamma)$, with
    \begin{align}
  \Sigma_\gamma = \Sigma_* + \sum\nolimits_{i=1}^{N_\Delta} \gamma_i \Sigma^\Delta_i,
  \label{eq:sigma_param}
\end{align}
where, following the notation of section~\ref{sec:model-description}, $\Sigma_*$ represents the nominal covariance, and $\Sigma_i^\Delta$ are several possible deviations.
The goal is to distinguish which normal distribution with covariance either $\Sigma_\alpha$ or $\Sigma_\beta$ (parameterized by $\alpha, \beta \in \mathbb{R}^{N_\Delta}$, resp.) better approximates $\mathcal{N}(0, \Sigma_\gamma)$.

To this end, we can compute the cumulative log-likelihood ratio of $\{y_k\}_{k=1}^N$ between the two covariance matrices as:
\begin{align}
  &\mathcal{L}(\{y_k\}_{k=1}^N|\alpha, \beta)
  =
  \sum_{k=1}^N \log \frac{
      \frac{1}{\sqrt{2\pi|\Sigma_\alpha|}}\exp\left[-y_k^\top \Sigma_\alpha^{-1} y_k\right]
    }{
      \frac{1}{\sqrt{2\pi|\Sigma_\beta|}}\exp\left[-y_k^\top \Sigma_\beta^{-1} y_k\right]
    } \nonumber
  \\ &\hspace{1cm} = \frac{1}{2} \left[N \log \frac{\left|\Sigma_\beta\right|}{\left|\Sigma_\alpha\right|} + \sum_{k=1}^N y_k^\top (\Sigma_\beta^{-1} - \Sigma_\alpha^{-1}) y_k\right]. \label{eq:llr-covariance}
\end{align}
When the ratio is positive, $\alpha$ is the more accurate model; otherwise, $\beta$.

\subsection{Statistical properties of the log-likelihood ratio}
We are now interested in characterizing the statistical properties of the log-likelihood ratio. The following results can then be used in practice to evaluate the detection mechanism's accuracy. We start with the following general characterization; see proof in the Appendix.

\begin{lemma}\label{lem:llr-distribution}
    The cumulative log-likelihood ratio between the distributions characterized by $\Sigma_\alpha$ and $\Sigma_\beta$ follows, when $N \to \infty$, the distribution
    $
      \mathcal{L}(\{y_k\}_{k=1}^N|\alpha, \beta) \sim \mathcal{N}(\mu, \sigma^2),
    $
    \begin{align}
      \mu &= \frac{N}{2} \left[
          \log \frac{\left|\Sigma_\beta\right|}{\left|\Sigma_\alpha\right|} + \mathrm{tr}\left((\Sigma_\beta^{-1} - \Sigma_\alpha^{-1}) \Sigma_\gamma\right)
        \right],
      \\
      \sigma^2 &= \frac{N}{2} \mathrm{tr}\left[(\Sigma_\beta^{-1} - \Sigma_\alpha^{-1}) \Sigma_\gamma (\Sigma_\beta^{-1} - \Sigma_\alpha^{-1}) \Sigma_\gamma\right].
    \end{align}
\end{lemma}

Note that, when $\Sigma_\gamma = \Sigma_\beta$, the expected value of the log-likelihood ratio becomes $N\cdot D_\mathrm{KL} \left(\mathcal{N}(0, \Sigma_\alpha), \mathcal{N}(0, \Sigma_\beta)\right)$, where $D_\mathrm{KL}$ is the Kullback-Leibler divergence. This connection is to be expected, as the KL divergence is the expected value of the log-likelihood ratio \cite{cover_elements_2006}.

With the characterization of its distribution in place, we can determine the probability of the cumulative log-likelihood being positive, that is, $\alpha$ being favored over $\beta$:
\begin{align}
    P(\mathcal{L}(\{y\}_{k=1}^N|\alpha, \beta) > 0) = \mathrm{cdf}\left(\frac{\mu}{\sigma}\right) = \int_{-\infty}^\frac{\mu}{\sigma} \frac{e^{-\frac{t^2}{2}}}{\sqrt{2 \pi}} d t, \label{eq:cdf}
\end{align}
where $\mathrm{cdf}(\cdot)$ is the cumulative distribution function of the normal distribution with mean $0$ and unit variance.

The formula~\eqref{eq:cdf} is difficult to interpret, as it does not make use of the linear combination structure~\eqref{eq:sigma_param} for $\Sigma_\gamma$ (and $\Sigma_\alpha$, $\Sigma_\beta$). The following lemma then simplifies it in the case of interest; see proof in the Appendix.

\begin{lemma}\label{lem:approx-distribution}
    Under the parameterization given by \eqref{eq:sigma_param}, the mean and covariance of the log-likelihood can be approximated quadratically around $\Sigma_*$ as
    \begin{align}
      \mu &= \frac{N}{4} (\alpha + \beta - 2 \gamma)^\top \Lambda (\alpha - \beta) + \mathcal{O}(\alpha^3, \beta^3, \gamma^3), \label{eq:mu_quad}
      \\
      \sigma^2 &= \frac{N}{2} (\alpha - \beta)^\top \Lambda (\alpha - \beta) + \mathcal{O}(\alpha^3, \beta^3, \gamma^3), \label{eq:sigma_quad}
    \end{align}
    where
    $
      \Lambda_{ij} = \mathrm{tr}\left[\Sigma_*^{-1} \Sigma_i^\Delta \Sigma_*^{-1} \Sigma_j^\Delta\right].
    $
\end{lemma}

Connecting again with the field of statistics, we note that $\Lambda$ corresponds to the Fisher information of the distribution $\mathcal{N}(0, \Sigma_*)$. This is again expected, as the Fisher information quantifies the amount of information a random variable carries about a parameter of its distribution \cite{cover_elements_2006}.

The analytical expressions \eqref{eq:mu_quad} and \eqref{eq:sigma_quad} provide insights into which deviations in the covariances are easier or harder to detect.
For instance, if $\beta$ is an accurate model of the actual system ($\beta = \gamma$), combining \eqref{eq:cdf}, \eqref{eq:mu_quad}, and \eqref{eq:sigma_quad} leads to
\begin{align*}
    &P\left(\mathcal{L}(\{y\}_{k=1}^N|\alpha, \beta\right)
    = \mathrm{cdf}\left(\frac{\mu}{\sigma}\right)
    \\ &\qquad \approx \mathrm{cdf}\left(- \frac{\sqrt{N}}{2 \sqrt{2}} \sqrt{(\alpha-\beta)^\top \Lambda (\alpha-\beta)}\right).
\end{align*}
Then, the probability of the wrong model ($\alpha$) obtaining a higher score falls at a rate of $\sqrt{(\alpha-\beta)^\top \Lambda (\alpha-\beta)}$; this quantifies the false positive probability.
Thus, if the models are similar, reliable model discrimination will require more measurements.
On the other hand, if $\gamma = (\alpha + \beta)/2$ (the hypotheses are equally accurate), the average value of the cumulative log-likelihood ratio is $0$, and, as expected, no hypothesis is favored.

More generally, as discussed previously, $\gamma$ is unknown in practice. We can then plug an estimated value $\hat{\gamma}$ into Lemma~\ref {lem:approx-distribution}, \textit{e.g.} constructed from historical data or first principles, to evaluate the accuracy of our detection.

\section{Step 2: Log-Likelihood-Based Anomaly Detection} \label{sec:dynamic}
We are now ready to turn our attention to the problem of interest: selecting the model, between $\Pi_\alpha$ and $\Pi_\beta$, that most accurately predicts the measurements sequence $\{ y_k \}_{k = 1}^N$ generated by~\eqref{eq:system_model} with $\Pi_\gamma$.
Following our discussion in section~\ref{sec:problem}, we are focused on systems with small deviations that require a long sequence of measurements for reliable detection. Thus, we assume throughout that most measurements are collected when the system has settled into steady-state.
To this end, the first step is to translate the results of section~\ref{sec:static} into suitable tools.

% We now proceed to analyze the case in which the measurements come from a dynamical system described by \eqref{eq:system_model}. Our approach is as follows: since detection times are expected to be long, most measurements will be obtained after the system has settled into steady state. Then, we can focus our analysis on the steady-state covariance of the measurements, $\Pi^y$. We will then determine how small variations in the matrices of the dynamical system ($A$, $C$, $Q$, and $R$) affect $\Pi^y$ and, with knowledge of the nominal $\Pi^y$ and the directions in which it can vary, we will use the results obtained for the static case.

\subsection{Detection via measurement covariance}
To leverage Section~\ref{sec:static}, we need to reformulate the detection problem as discriminating between covariance matrices that better model the sequence of measurements.

\subsubsection{Nominal measurement covariance}
For a nominal plant model $\Pi_*$, the covariance of $x_k$ at steady-state $\mathbb{E}[x_k x_k^\top] = \Sigma_*^x$ can be obtained based on the fact that it does not depend on $k$, which leads to the discrete Lyapunov equation:
\begin{align}
  \Sigma_*^x &= \mathbb{E}[x_{k+1} x_{k+1}^\top]
  = \mathbb{E}\left[(A^* x_k + w_k)(A^* x_k + w_k)^\top\right] \nonumber
  \\ &= A_* \mathbb{E}\left[x_k x_k^\top\right] A_*^\top + \mathbb{E}[w_k w_k^\top] 
  = A_* \Sigma_*^x A_*^\top + Q_*. \label{eq:cov_x_nom}
\end{align}
This equation can be solved easily either with the direct or the bilinear method. After computing the steady state covariance of $x$, the covariance of the measurements is then characterized by:
\begin{align}
    &\Sigma_*^y = \mathbb{E}[y_k y_k^\top] 
    = \mathbb{E}\left[\left(C_* x_k + v_k\right) \left(C_* x_k + v_k\right)^\top \right] \nonumber
    \\ &= C_* \mathbb{E}\left[x_k x_k^\top\right] C_*^\top + \mathbb{E}\left[v_k v_k^\top \right]
    = C_* \Sigma_*^x C_*^\top + R_*. \label{eq:cov_y_nom}
\end{align}

\subsubsection{Deviation of measurement covariance}
Given the measurement covariance for the nominal model $\Pi_*$, we now want to determine how it changes when we make small changes to the model, described by the model deviations $\Pi_i^\Delta$. For a model deviation $\Pi_i^\Delta$, the steady-state covariance of the state of $\Pi_* + \gamma_i \Pi_i^\Delta$ can be computed as
\begin{align*}
    \Sigma_*^x + \gamma_i \Sigma_i^{x, \Delta} &= (A_* + \gamma_i A_i^\Delta) (\Sigma_* + \gamma_i \Sigma_i^{x, \Delta}) (A_* + \gamma_i A_i^\Delta)^\top \nonumber
    \\ &+ Q_* + \gamma_i Q_i^\Delta.
\end{align*}
We then linearize this expression with respect to $\gamma$ by removing higher-order terms, which yields the discrete Lyapunov equation:
\begin{equation}
    \Sigma_i^{x, \Delta}
    =
    A_* \Sigma_i^{x, \Delta} A_*^\top
    +
    A_i^\Delta \Sigma_* A_*^\top + A_* \Sigma_* {A_i^\Delta}^\top + Q_i^\Delta.
    \label{eq:cov_x_dev}
\end{equation}
We can follow a similar procedure to obtain $\Sigma_i^{y,\Delta}$:
\begin{align*}
    &\Sigma_*^y + \gamma_i \Sigma^{y, \Delta}
    = \nonumber
    \\ &(C_* + \gamma_i C^\Delta) (\Sigma_*^x + \gamma_i \Sigma^{x, \Delta}) (C_* + \gamma_i C^\Delta)^\top \nonumber
     + R_* + \gamma_i R_i^\Delta,
\end{align*}
and removing higher order terms again, we obtain
\begin{equation}
    \Sigma_i^{y, \Delta}
    =
    C_* \Sigma_i^{x, \Delta} C_*^\top + C_* \Sigma_*^{x} {C_i^\Delta}^\top + C_i^\Delta \Sigma_*^x C_*^\top + R_i^\Delta.
    \label{eq:cov_y_dev}
\end{equation}

Then, given a nominal plant $\Pi^*$ and a set of possible deviations $\Pi_i^\Delta$, $i \in \{ 1, \ldots, N_\Delta \}$, we can compute the nominal measurement covariance $\Pi_*^y$ from \eqref{eq:cov_x_nom} and \eqref{eq:cov_y_nom}, and the variation of the covariance associated with each deviation using \eqref{eq:cov_x_dev} and \eqref{eq:cov_y_dev}. For a system $\Pi_\gamma$, its steady state measurement covariance matrix $\Sigma_\gamma^y$ can therefore be approximated as (recalling that $\| \gamma \| \ll 1$):
\begin{equation}\label{eq:sigma_gamma}
    \Sigma_\gamma^y \approx \Sigma_*^y + \sum_{i=1}^{N_\Delta} \gamma_i \Sigma_i^{y, \Delta}.
\end{equation}

\subsubsection{Detection algorithm}\label{par:detection-algorithm}
With this derivation in place, we can then design the following detection algorithm:
\begin{enumerate}
\setcounter{enumi}{-1}
    \item Input: measurement sequence $\{ y_k \}_{k=1}^N$, nominal model $\Pi_*$, hypothesis models $\Pi_\alpha$ and $\Pi_\beta$, with $\alpha, \beta \in \mathbb{R}^{N_\Delta}$.

    \item Compute the nominal steady-state covariance of the measurements $\Sigma_*^y$ (see~\eqref{eq:cov_y_nom}), and for each deviation $i$, the variation matrix $\Sigma^{y,\Delta}_i$ (see~\eqref{eq:cov_y_dev}).

    \item Compute $\Sigma_\alpha^y$ and $\Sigma_\beta^y$ applying~\eqref{eq:sigma_gamma} to $\Pi_\alpha$ and $\Pi_\beta$.

    \item Finally, compute the cumulative log-likelihood ratio according to~\eqref{eq:llr-covariance}. Then select the hypothesis model $\alpha$ if the ratio is positive, and $\beta$ otherwise.
\end{enumerate}

\subsection{Evaluating detection accuracy}

\subsubsection{Detection accuracy}\label{subsec:detection-accuracy}
Recall from section~\ref{sec:problem} that the goal is to select the hypothesis model, between $\Pi_\alpha$ and $\Pi_\beta$, which more accurately approximates the actual model $\Pi_\gamma$.
Leveraging the above derivation, we can thus effectively reformulate this problem as the problem of selecting the more accurate approximation of the output covariance matrix $\Sigma_\gamma^y$ in~\eqref{eq:sigma_gamma}.
Indeed, applying~\eqref{eq:sigma_gamma} to $\Pi_\alpha$ and $\Pi_\beta$ yields the hypothesis covariance matrices $\Sigma_\alpha^y$ and $\Sigma_\beta^y$, and by Lemma~\ref{lem:approx-distribution} the probability of $\alpha$ being favored over $\beta$ is determined by:
\begin{align}
    &P(\mathcal{L}(\{y_k\}_{k=1}^N|\alpha, \beta) > 0) \nonumber
    \\ &\qquad \approx \mathrm{cdf}\left(
        \frac{-\sqrt{N}}{2\sqrt{2}}
        \frac{
            (\alpha + \beta - 2\gamma)^\top \Lambda (\alpha - \beta)
        }{
            (\alpha - \beta)^\top \Lambda (\alpha - \beta)
        }
    \right), \label{eq:probability-false-positive}
\end{align}
where $\Lambda_{ij} = \mathrm{tr}\left[(\Sigma_*^y)^{-1} \Sigma_i^{y,\Delta} (\Sigma_*^y)^{-1} \Sigma_j^{y,\Delta}\right]$.

\subsubsection{Correcting for measurement correlation}\label{sec:correlation}
A limitation in our approach is that it relies on Lemma~\ref{lem:approx-distribution}, which was derived under the assumption of i.i.d. measurements $\{ y_k \}_{k=1}^N$.
However, the outputs of a dynamical system are not uncorrelated, as shown by the covariance matrix:
\begin{align}
    &\mathbb{E}\left[y_{k+t} y_k^\top\right]
    = \mathbb{E}\left[
        \left(
            C x_{k+t} + v_{k}
        \right)\left(
            C x_k + v_k
        \right)^\top
    \right] \nonumber
    \\
    &= \mathbb{E}\left[
        \left(
            C \left(A^t x_{k} + \sum_{i=k}^{k+t} A^{i-k} w_k\right)+ v_{k}
        \right)\left(
            C x_k + v_k
        \right)^\top
    \right] \nonumber
    \\ &= C A^t \Sigma^x C^\top. \label{eq:correlation}
\end{align}
It is therefore important to assess how this correlation impacts the cumulative log-likelihood ratio as a detection mechanism. In particular, we first characterize this deviation when $y_k \in \mathbb{R}$ in Lemma~\ref{lem:correction} (proved in Appendix), and then translate it into a heuristic to correct $P(\mathcal{L}(\{y_k\}_{k=1}^N|\alpha, \beta) > 0)$ when $y_k \in \mathbb{R}^p$, $p \geq 1$.

\begin{lemma}\label{lem:correction}
    Let $\{ y_k \}_{k=1}^N$, $y_k \in \mathbb{R}$, be output measurements of~\eqref{eq:system_model} such that $\mathbb{E}[y_k y_{k+t}] = \lambda^t \mathbb{E}[y_k y_k]$, with $\lambda \in [0, 1)$.
    Then, as $N \to \infty$, the distribution of the cumulative log-likelihood ratio approximates a normal distribution with mean $\mu$ and variance $\sigma^2$ defined as
    \[
        \mu = \mu_\mathrm{uncorr.}, \quad \sigma^2 = \sigma^2_\mathrm{uncorr} \frac{1+\lambda^2}{1-\lambda^2}, 
    \]
    where $\mu_\mathrm{uncorr.}$ and $\sigma^2_\mathrm{uncorr.}$ are the mean and variance of the uncorrelated case, defined in Lemma~\ref{lem:llr-distribution}.
\end{lemma}

We can then take inspiration from this result to propose the following correction heuristic: the covariance between steps, given by~\eqref{eq:correlation}, suggests that it decays at a rate determined by the largest eigenvalue $\lambda_\mathrm{max}$ of $A$.
Therefore, we correct our approximation by:
$$
    P_\mathrm{corr} (\mathcal{L} > 0) = \mathrm{cdf}\left(\frac{\mu}{\sigma}\sqrt{\frac{1 - \lambda_\mathrm{max}^2}{1 + \lambda_\mathrm{max}^2}}\right).
$$
We can then leverage this approximation to provide a confidence bound on the cumulative log-likelihood ratio, indicating that a stronger correlation between measurements yields a less accurate detection threshold.

\section{Application to Detection and Observer Design} \label{sec:experiments}
In this section, we illustrate the application of the proposed results to: 1) detection of deviations from the nominal model, and 2) observer design to increase detection sensitivity. \footnote{The code can be found in \url{https://github.com/AlejandroPenacho/cdc26-log-likelihood}}

\subsection{Deviation detection for double pendulum}\label{subsec:double-pendulum}
\paragraph*{Setup}
We consider an inverted double pendulum, stabilized in vertical position by state feedback which controls the torque at the two joints. Let $\theta_1$ and $\theta_2$ be the angles of the first and second joint (w.r.t. the vertical). Then the state vector is $x = \begin{bmatrix} \theta_1 & \dot{\theta}_1 & \theta_2 & \dot{\theta}_2 \end{bmatrix}$, and linearizing and discretizing yields the open loop model:
\begin{align*}
    A_o &= \begin{bmatrix}
        1 & T_s & 0 & 0
        \\
        \frac{T_s m_t g}{m_1 l_1} & 1-\frac{T_s d_1}{m_1 l_1^2} & -\frac{T_s m_2 g}{m_1 l_1} & \frac{T_sd_2}{m_1 l_1 l_2}
        \\
        0 & 0 & 1 & T_s
        \\
        - \frac{T_s m_t g}{m_1 l_2} & \frac{T_s d_1}{m_1 l_1 l_2} & \frac{T_s m_t g}{m_1 l_2} & 1-\frac{T_s m_t d_2}{m_1 m_2 l_2^2}
    \end{bmatrix},
    \\
    B &= T_s\begin{bmatrix}
        0 & 0
        \\
        \frac{1}{m_1 l_1^2} & -\frac{1}{m_1 l_1 l_2}
        \\
        0 & 0
        \\
        -\frac{1}{m_1 l_1 l_2} & \frac{m_t}{m_1 m_2 l_2^2}
    \end{bmatrix},
    \qquad
    C = \begin{bmatrix}
        1 & 0 & 0 & 0
        \\
        0 & 0 & 1 & 0
    \end{bmatrix},
    \\
    Q &= \mathrm{diag}(0, 0.5, 0, 0.6), \qquad \qquad \quad R = 0.1 I.
\end{align*}
The parameters are the masses of the pendulums $m_1=1$ and $m_2=1.2$ kg, with $m_t = m_1 + m_2$ kg, the friction at the joints $d_1=d_2=0.3\ \mathrm{Nms}$, gravity acceleration $g=9.8\ \mathrm{m/s}^2$, lengths of the arms $l_1=0.5$ and $l_2=0.2$, and sampling time $T_s=0.2$ s.
The state feedback gain is:
$
    K = \begin{bmatrix}
        33.04 & 6.76 & 1.33 & 0.99 \\
        4.72 & 1.52 & 3.54 & 0.2
    \end{bmatrix},
$
which places the closed-loop poles at $-5\times10^{-4}$, $-0.6$, $-0.2$ and $-0.4$.
In this scenario, the nominal model of the closed-loop plant is $\Pi_* = (A_o - B K, C, Q, R)$.

We consider three possible deviations: friction increase at the first or second joint (modifying either $d_1$ or $d_2$), and an increase in torque noise at the first joint, modeled as:
\begin{align*}
    A_1^\Delta = \frac{\partial}{\partial d_1} A,
    \quad
    A_2^\Delta = \frac{\partial}{\partial d_2} A,
    \quad
    Q_3^\Delta = \mathrm{diag}(0, 0.5, 0, 0),
\end{align*}
where the partial derivatives are evaluated for the numerical values of $m_1, m_2, d_1, d_2, g, l_1, l_2$ introduced above, and all unspecified deviation matrices are set to $0$.

\paragraph*{Detection sensitivity}
Applying the results in section~\ref{subsec:detection-accuracy}, we characterize the matrix $\Lambda$ and its correlation normalization ($R$, $D$) as
\begin{align*}
    \Lambda &= \begin{bmatrix}
        1.1 & 7.09 & 1.25\\
        7.09 & 72.69 & 24.33\\
        1.25 & 24.33 & 13.25  
    \end{bmatrix} = D^\frac{1}{2} R D^\frac{1}{2}
    \nonumber
    \\ R &= \begin{bmatrix}
        1.00 & 0.80 & 0.33\\
        0.80 & 1.00 & 0.78\\
        0.33 & 0.78 & 1.00  
    \end{bmatrix}
    \quad D =
    \begin{bmatrix}
        1.1 &  0 & 0\\
        0 & 72.76 & 0\\
        0 & 0 & 13.25
    \end{bmatrix}.
\end{align*}
We can now employ these matrices to characterize the detectability of different deviations.
First of all, a larger element on the diagonal of $\Lambda$ indicates a deviation that is easier to detect (in this case, friction increase on the second joint is the easiest to detect).
Secondly, the correlation matrix $R$ provides a picture of how deviations interact with each other. A large off-diagonal element $R_{ij}$ signifies that deviations $i$ and $j$ have a similar impact on measurements, and are thus more difficult to distinguish.
Generally, small off-diagonals are preferable for more accurate classification of deviations. Nonetheless, the ability to detect that a deviation has occurred (even without correctly classifying) is still of great value.

\begin{figure}[!ht]
    \centering
    \includegraphics[width=\linewidth]{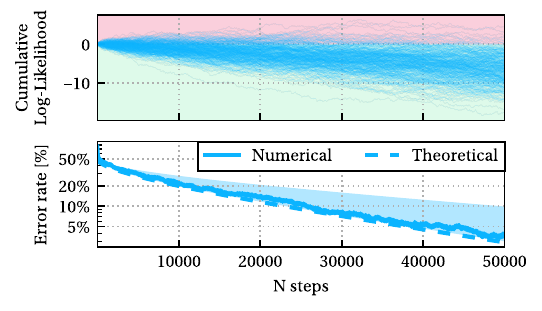}
    \caption{\textbf{Upper:} $2000$ trajectories of the cumulative log-likelihood for the stabilized inverted double pendulums. \textbf{Lower:} empirical and theoretical error rates, with shaded confidence bounds.}
    \label{fig:example_pendulum}
\end{figure}

\paragraph*{Numerical evaluation}
We consider now $\gamma = [0.03 \quad 0. \quad 0.01]$ (which is not available to the detection algorithm), and candidate models, $\alpha = [0.\quad 0.\quad 0.]$ and $\beta = [0.04 \quad 0 \quad 0]$, neither of them tuned to the actual system, but $\beta$ closer to it.
In this setup, we evaluate the performance of the detection algorithm of section~\ref{par:detection-algorithm} as follows.
We run $2000$ simulations during which we collect $N = 50000$ measurements from the system, and use~\eqref{eq:llr-covariance} to compute the cumulative log-likelihood ratio at each measurement time $k$. The evolution of the ratio is plotted in Fig.~\ref{fig:example_pendulum} (upper).
At each time $k$, we then compute the percentage of trajectories for which the ratio is positive (red region), and plot it in Fig.~\ref{fig:example_pendulum} (solid line in the lower plot).
As we can see, the percentage decreases from $\sim 50\%$ ($\alpha$ and $\beta$ are equally accurate), reaching below $5\%$ as more measurements are collected. This indicates that the algorithm successfully selects $\beta$ as the better approximation of $\gamma$.

As discussed in section~\ref{sec:correlation}, the correlation of measurements for a dynamical system will influence the accuracy of the log-likelihood ratio.
Using the ground truth $\gamma$, we can now compute the theoretical probability of the log-likelihood being positive as
$
   P(\mathcal{L} > 0) = \mathrm{cdf}(-0.00846 \sqrt{N} )
$
and its corrected version (with $\lambda_\mathrm{max} = 0.6$) as
$
   P_\mathrm{corr} (\mathcal{L} > 0) = \mathrm{cdf}(-0.012\sqrt{N} ).
$
We then plot in Fig.~\ref{fig:example_pendulum} the theoretical error rate and the confidence bound given by the corrected probability.
We see that the numerical decay-rate is slightly larger than the theoretical one, but well within the confidence bounds, demonstrating the effectiveness of this detection mechanism.

% \begin{align*}
%     P(\mathcal{L} > 0) &= \mathrm{cdf}\left(
%     -\frac{\sqrt{N}}{2 \sqrt{2}} \frac{
%         (\alpha - \beta - 2\gamma)^\top \Lambda (\alpha - \beta)
%     }{
%         (\alpha - \beta)^\top \Lambda (\alpha - \beta)
%     }
%     \right) \nonumber
%     \\ &= \mathrm{cdf}\left(-0.00846 \sqrt{N} \right)
%     \\
%     P_\mathrm{corr} (\mathcal{L} > 0) &= \mathrm{cdf}\left(-0.0845 \sqrt{\frac{1+\lambda_\mathrm{max}^2}{1-\lambda_\mathrm{max}^2}} \sqrt{N} \right)
%     \nonumber
%     \\
%     &= \mathrm{cdf}\left(-0.012\sqrt{N} \right)
% \end{align*}

% \noindent\rule{5.15in}{0.05in}
\subsection{Observer design for friction system}

\paragraph*{Setup}
In this section, we investigate how the log-likelihood ratio can be leveraged to design observers that maximize the detectability of deviations.
We consider a friction system stabilized around an operating point, with state space dynamics described by
\begin{align*}
    A &= \begin{bmatrix}
        T_s(-f_v - K_p) & -T_sK_i \\
        T_s & 1
    \end{bmatrix},
    \quad & Q &= \begin{bmatrix}0.1 & 0 \\ 0 & 0 \end{bmatrix},
    \\
    C &= \begin{bmatrix}1 & 0\end{bmatrix}, \quad &R&=\begin{bmatrix} 0.5 \end{bmatrix},
\end{align*}
with $T_s = 0.2\ \mathrm{s}$, $f_v = 2.0\ \mathrm{Nms}$, $K_p = 5.0\ \mathrm{Nms}$ and $K_i = 2.5\ \mathrm{Nm}$.
We then include an observer ($(\hat{L}, \hat{C})$, to be designed) by augmenting the state to $\bar{x} = \begin{bmatrix} x^\top & \hat{x}^\top & v^\top \end{bmatrix}$, which contains the estimated state $\hat{x}$ and the observation noise $v$, treated as process noise; the resulting system is:
\begin{align*}
    \bar{A} &= \begin{bmatrix}
        A & 0 & 0
        \\
        -\hat{L} \hat{C} & \hat{L} \hat{C} & I
        \\
        0 & 0 & 0
    \end{bmatrix},
    &
    \bar{Q} &= \begin{bmatrix}
        Q & 0 & 0 \\ 0 & 0 & 0 \\ 0 & 0 & R
    \end{bmatrix},
    \\
    \bar{C} &= \begin{bmatrix}
        C & -\hat{C} & I
    \end{bmatrix},
    &
    \bar{R} &= 0.
\end{align*}
Note that the new output of the system is the innovation of the observer $y_k - \hat{C} \hat{x}_k$.
Finally, we consider one deviation: an increase in viscous friction, defined as $\bar{A}_i^\Delta = \partial \bar{A}/\partial f_v$.

% For the new system, deviations in the model are different depending on whether they affect the observer (matrices $\hat{A}, \hat{L}, \hat{C}$) or the actual plant ($A$, $C$, $Q$, $R$). For the nominal model $\Pi_*$, it is reasonable to make the equal. But when considering a model deviation for the plant, it is important to keep in mind that it would only affect the actual plant $A$, and not the observer.

\paragraph*{Observer design}
\begin{figure}[!ht]
    \centering
    \includegraphics[width=\linewidth]{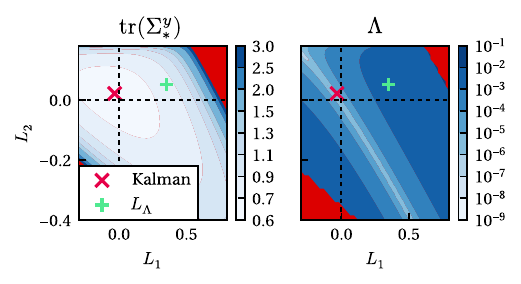}
    \caption{Contours of $\mathrm{tr}(\Sigma_*^y)$ and $\Lambda$ for different gains of the observer $L$. Red areas show unstable observer configuration.}
    \label{fig:contours}
\end{figure}
The goal now is to leverage the log-likelihood ratio to select an observer that maximizes detectability.
To this end, for a range of possible observer gains $L = [L_1, L_2]$, we compute: 1) matrix $\Lambda$ characterizing detectability, and 2) $\mathrm{tr}(\Sigma^y)$, which represents the average squared innovation.
We plot the resulting values in Fig.~\ref{fig:contours} as a heatmap in the $(L_1, L_2)$-plane.
As expected, the Kalman filter gain minimizes $\mathrm{tr}(\Sigma^y)$ (left plot). However, the Kalman gain fails to maximize detectability (right plot), which instead is achieved by the gain $L_\Lambda = [0.35, 0.05]$.
On the other hand, $L_\Lambda$ attains a value of $\mathrm{tr}(\Sigma^y)$ that is twice as large as that of the Kalman gain.
This highlights an important trade-off between the observer's accuracy and its detectability.

\paragraph*{Numerical evaluation}
\begin{figure}[!ht]
    \centering
    \includegraphics[width=\linewidth]{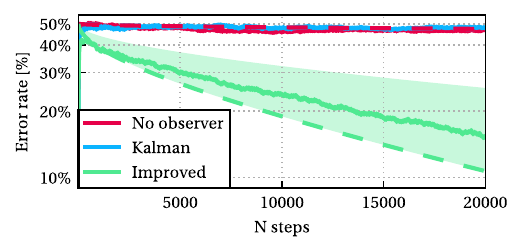}
    \caption{Experimental (solid line) and theoretical (dashed line) error rates with 3 different observer configurations.}
    \label{fig:example_friction}
\end{figure}
We now compare the detectability when employing no observer ($L = [0, 0]$), the Kalman filter, and $L_\Lambda$.
Specifically, we consider $\gamma = 0.5$ (the weight on the friction deviation), and the hypothesis models $\alpha = 0$ and $\beta=\gamma$. Following the same procedure as in section~\ref{subsec:double-pendulum}, we plot in Fig.~\ref{fig:example_friction} the numerical and theoretical error rates for all three gain choices, along with their respective confidence bounds.
We can see that $L_\Lambda$, chosen to maximize $\Lambda$, indeed significantly improves detectability, while the detectability with the Kalman gain (or without observer) is quite poor.
However, the Kalman gain reduces the gap between the theoretical and numerical error rates. This is because the Kalman filter makes innovation measurements uncorrelated, and, as shown in section~\ref{sec:correlation}, this implies that Lemma~\ref{lem:approx-distribution} provides a good approximation of the empirical error rate.
This again highlights the need to find the right balance between observer accuracy and detectability.

%%%%%%%%%%%%%%%%%%%%%%%%%%%%%%%%%%%%%%%%%%%%%%%%%%%%%%%%%%%%%%%%%%%%%%%%%%%%%%%%
\section{Conclusions}\label{sec:conclusions}
In this paper, we have investigated a detection mechanism for dynamical systems subject to small deviations from the nominal behavior.
This method relies on the cumulative log-likelihood ratio, which allows us to theoretically characterize the detector's error rate. We have then showcased the promise of this approach with applications to detection and observer design.
This work opens several directions for future research. These include extending to non-autonomous systems and leveraging input for more accurate detection, and exploring our observer design approach for larger-scale systems.

\section*{Usage Of Generative AI}
ChatGPT (OpenAI) was used to assist with language refinement. Grammarly was used for grammar and spelling checks. All content was reviewed and verified by the authors.

\appendix

\subsection{Proof of Lemma 1}\label{sec:proof_a}
The cumulative log-likelihood ratio between two zero-mean normal distributions $\Sigma_\alpha$ and $\Sigma_\beta$ for a sequence of i.i.d. observations distributed each as $y_k \sim \mathcal{N}(0, \Sigma_\gamma)$ is given by:
\begin{align}
    \mathcal{L}(\{y_k\}_{k=1}^N | \Sigma_\alpha, \Sigma_\beta)
    &=
    \frac{1}{2} \left[ N\log \frac{|\Sigma_\beta|}{|\Sigma_\alpha|} \right.
    \nonumber
    \\
    &\left. + \sum_{k=1}^N y_k^\top (\Sigma_\beta^{-1} - \Sigma_\alpha^{-1}) y_k \right] \label{eq:llr-derive}
\end{align}

The first component is constant, so we focus on the second. We first define $F = \Sigma_\gamma^\frac{1}{2} \left(\Sigma_\beta^{-1} - \Sigma_\alpha^{-1}\right)\Sigma_\gamma^\frac{1}{2}$, and since $F = F^\top$, we can diagonalize it with the eigenvalue decomposition $F = U S U^\top$, where $S = \mathrm{diag}(\lambda_1, \lambda_2, \dots)$, $\lambda_i$ being the $i$-th eigenvalue of $F$. We can then define the variable $z_k = U^\top \Sigma_\gamma^{-\frac{1}{2}} y_k$, with covariance given by $\mathbb{E}[z_k z_k^\top] = I$. We use these two definitions to rewrite the second component of \eqref{eq:llr-derive} as

\begin{align*}
    &\frac{1}{2} \sum_{k=1}^N y_k^\top \left(\Sigma_\beta^{-1} - \Sigma_\alpha^{-1}\right) y_k
    =
    \frac{1}{2} \sum_{k=1}^N z_k^\top S z_k
    \\ =
    &\frac{1}{2} \sum_{k=1}^N \sum_{d=1}^n \lambda_d z_{k,d}^2,
    =
    \frac{1}{2} \sum_{d=1}^n \lambda_d \sum_{k=1}^N z_{k, d}^2,
\end{align*}
where we have used the fact that $S$ is diagonal to write the result as a sum. For the next step, we use the property that the values of $z_{k,d}$ are uncorrelated to obtain a $\chi^2$ distribution with $N$ degrees of freedom, and due to the assumption of $N$ being large, we approximate the distribution as normal:
\begin{align*}
    &\frac{1}{2} \sum_{d=1}^n \lambda_d \sum_{k=1}^N z_{k, d}^2 \nonumber
    \sim \frac{1}{2} \sum_{d=1}^n \lambda_d \chi_N^2
    \approx \sum_{d=1}^n \frac{\lambda_d}{2} \mathcal{N}(N, 2N) \nonumber
    \\ &= \sum_{d=1}^n \mathcal{N}\left( \frac{N}{2} \lambda_d, \frac{N}{2} \lambda_d^2\right) \nonumber
    = \mathcal{N}\left( \frac{N}{2} \sum_{d=1}^n \lambda_d, \frac{N}{2} \sum_{d=1}^n \lambda_d^2\right).
\end{align*}

To compute the sum of the eigenvalues and the squared eigenvalues of $F$, we use the property that these are equal to the traces of $F$ and $F^2$, respectively, and simplifying the expression with the property $\mathrm{tr}(AB) = \mathrm{tr}(BA)$:

\begin{align*}
    &\sum_{d=1}^n \lambda_d = \mathrm{tr}\left[F\right]
    = \mathrm{tr}\left[ \Sigma_\gamma^{\frac{1}{2}} \left(\Sigma_\beta^{-1} - \Sigma_\alpha^{-1} \right) \Sigma_\gamma^{\frac{1}{2}}\right]
    \\ &= \mathrm{tr}\left[\left(\Sigma_\beta^{-1} - \Sigma_\alpha^{-1} \right) \Sigma_\gamma\right]
    \\
    &\sum_{d=1}^n \sigma_d^2(F) = \mathrm{tr}\left[F F\right]
    \\ &= \mathrm{tr}\left[
        \Sigma_\gamma^{\frac{1}{2}} \left(\Sigma_\beta^{-1} - \Sigma_\alpha^{-1} \right) \Sigma_\gamma^{\frac{1}{2}}
        \Sigma_\gamma^{\frac{1}{2}} \left(\Sigma_\beta^{-1} - \Sigma_\alpha^{-1} \right) \Sigma_\gamma^{\frac{1}{2}}
    \right]
    \\ &= \mathrm{tr}\left[
        \left(\Sigma_\beta^{-1} - \Sigma_\alpha^{-1} \right) 
        \Sigma_\gamma \left(\Sigma_\beta^{-1} - \Sigma_\alpha^{-1} \right) \Sigma_\gamma
    \right]
\end{align*}

Adding to the mean the first component of \eqref{eq:llr-derive}, we obtain the expression shown in the lemma.

\subsection{Proof of Lemma 2}\label{sec:proof_b}
We can compute the derivatives of the mean and the variance using the standard rules of derivation for matrices, namely:
\begin{align*}
    \frac{\partial \log |\Sigma|}{\partial x} = \mathrm{tr}\left(\Sigma^{-1} \frac{\partial\Sigma}{\partial x}\right),
     \qquad
     \frac{\partial \Sigma^{-1}}{\partial x} = \Sigma^{-1} \frac{\partial \Sigma}{\partial x} \Sigma^{-1},
\end{align*}
and from our parameterization of the matrices considered, we have
\begin{align*}
    \frac{\partial\Sigma_\alpha}{\partial \alpha_i} = 
    \frac{\partial\Sigma_\beta}{\partial \beta_i} = 
    \frac{\partial\Sigma_\gamma}{\partial \gamma_i} = \Sigma_i^\Delta \quad \forall i.
\end{align*}

Following these standard derivation rules, the first and second order derivatives of $\mu$ and $\sigma^2$ can be obtained and evaluated at $\Sigma_\alpha = \Sigma_\beta = \Sigma_\gamma = \Sigma_*$:
\begin{align*}
    \frac{\partial \mu}{\partial \alpha_i}\Bigg|_* &= \frac{\partial \mu}{\partial \beta_i}\Bigg|_* = \frac{\partial \mu}{\partial \gamma_i}\Bigg|_* = 0
    \\
    \frac{\partial^2 \mu}{\partial \alpha_i \partial \alpha_j}\Bigg|_* &= \frac{\partial^2 \mu}{\partial \beta_i \partial \gamma_j}\Bigg|_* = \frac{N}{2} \mathrm{tr}\left[\Sigma_*^{-1} \Sigma_i^\Delta \Sigma_*^{-1} \Sigma_j^\Delta \right]
    \\
    \frac{\partial^2 \mu}{\partial \beta_i \partial \beta_j}\Bigg|_* &= \frac{\partial^2 \mu}{\partial \alpha_i \partial \gamma_j}\Bigg|_* = -\frac{N}{2} \mathrm{tr}\left[\Sigma_*^{-1} \Sigma_i^\Delta \Sigma_*^{-1} \Sigma_j^\Delta \right]
    \\
    \frac{\partial^2 \mu}{\partial \gamma_i \partial \gamma_j}\Bigg|_* &= \frac{\partial^2 \mu}{\partial \alpha_i \partial \beta_j}\Bigg|_* = 0
\end{align*}
Defining $\Lambda$ as a matrix where each entry is $\Lambda_{ij} = \mathrm{tr}(\Sigma_*^{-1} \Sigma_i^\Delta \Sigma_*^{-1} \Sigma_j^\Delta)$, the resulting quadratic expansion can be written as
\begin{align*}
    \mu &\approx \frac{1}{2} \frac{N}{2} \begin{bmatrix}
        \alpha^\top & \beta^\top & \gamma^\top
    \end{bmatrix} \begin{bmatrix}
        \Lambda & 0 & -\Lambda
        \\
        0 & -\Lambda & \Lambda
        \\
        -\Lambda & \Lambda & 0
    \end{bmatrix} \begin{bmatrix} \alpha \\ \beta \\ \gamma\end{bmatrix}
    \nonumber
    \\ &= \frac{N}{4} (\alpha + \beta - 2 \gamma)^\top \Lambda (\alpha - \beta)
\end{align*}

For the variances, the same process is followed to obtain
\begin{align*}
    \frac{\partial \sigma^2}{\partial \alpha_i}\Bigg|_* &= \frac{\partial \sigma^2}{\partial \beta_i}\Bigg|_* = \frac{\partial \sigma^2}{\partial \gamma_i}\Bigg|_* = 0
    \\
    \frac{\partial^2 \sigma^2}{\partial \alpha_i \partial \alpha_j}\Bigg|_* &= \frac{\partial^2 \sigma^2}{\partial \beta_i \partial \beta_j}\Bigg|_* = N \mathrm{tr}\left[\Sigma_*^{-1} \Sigma_i^\Delta \Sigma_*^{-1} \Sigma_j^\Delta \right]
    \\
    \frac{\partial^2 \sigma^2}{\partial \alpha_i \partial \beta_j}\Bigg|_* &= -N \mathrm{tr}\left[\Sigma_*^{-1} \Sigma_i^\Delta \Sigma_*^{-1} \Sigma_j^\Delta \right]
    \\
    \frac{\partial^2 \sigma^2}{\partial \alpha_i \partial \mu_j}\Bigg|_* &= \frac{\partial^2 \sigma^2}{\partial \beta_i \partial \mu_j}\Bigg|_* = \frac{\partial^2 \sigma^2}{\partial \mu_i \partial \mu_j}\Bigg|_* = 0
\end{align*}
The resulting quadratic expansion is then
\begin{align*}
    \sigma^2 &\approx \frac{1}{2} N \begin{bmatrix}
        \alpha^\top & \beta^\top & \gamma^\top
    \end{bmatrix} \begin{bmatrix}
        \Lambda & -\Lambda & 0
        \\
        -\Lambda & \Lambda & 0
        \\
        0 & 0 & 0
    \end{bmatrix} \begin{bmatrix} \alpha \\ \beta \\ \gamma\end{bmatrix}
    \nonumber
    \\ &= \frac{N}{2} (\alpha - \beta)^\top \Lambda (\alpha - \beta)
\end{align*}

\subsection{Proof of Lemma 3}\label{sec:proof_c}
We consider a sequence of measurements $y_k \in \mathbb{R}$. Each measurement has zero mean and variance $\sigma_y^2$, and measurements at different steps are correlated by a factor $\lambda \in [0, 1)$ that decays with the number of steps between them. The vector of all measurements $\bar{y}$ is then defined and distributed as
\begin{align*}
    \bar{y} &= \begin{bmatrix} y_1 \\ y_2 \\ \vdots\end{bmatrix} = \mathcal{N}(0, \bar{\Sigma}_y),
    \qquad
    \bar{\Sigma}_y = \sigma_y^2 \begin{bmatrix}
        1 & \lambda & \lambda^2 & \cdots \\
        \lambda & 1 & \lambda & \cdots \\
        \lambda^2 & \lambda & 1 & \cdots \\
        \vdots & \vdots & \vdots & \ddots
    \end{bmatrix}.
\end{align*}
We consider again the use of the cumulative log-likelihood ratio with two candidate variances $\sigma_\alpha^2$ and $\sigma_\beta^2$. To estimate its distribution after $N$ samples, we follow the same approach as for the uncorrelated case:
\begin{align*}
    \mathcal{L}(\{y_k\}_{k=1}^N|\sigma_\alpha^2, \sigma_\beta^2)
    = \frac{1}{2} \left[N \log \frac{\sigma_\beta^2}{\sigma_\alpha^2} + \sum_{k=1}^N (\sigma_\beta^{-2} - \sigma_\alpha^{-2})y_k^2 \right]
\end{align*}

Analogous to the proof of Lemma 1, we obtain the eigenvalue decomposition $\bar{\Sigma}_y = U S U^\top$, with $S=\mathrm{diag}(\lambda_1, \lambda_2, \cdots)$ and $\lambda$ being the eigenvalues of $\bar{\Sigma}_y$. Then we define $\bar{z} = U^\top \bar{\Sigma}_y^{-\frac{1}{2}} \bar{y}$, so it fulfills $\mathbb{E}[\bar{z}\bar{z}^\top]=I$, that is, the terms are uncorrelated. We can then reduce the second term of the previous formula, exploiting the fact that $S$ is diagonal:
\begin{align*}
    \sum_{k=1}^N (\sigma_\beta^{-2} - \sigma_\alpha^{-2}) y_k^2
    &= (\sigma_\beta^2 - \sigma_\alpha^2)\bar{y} \bar{y}^\top
    \nonumber
    \\ = (\sigma_\beta^2 - \sigma_\alpha^2) \bar{z} S \bar{z}^\top
    &= (\sigma_\beta^2 - \sigma_\alpha^2) \sum_{k=1}^N \lambda_k z_k^2
\end{align*}
Each term has mean $\lambda_k$, variance $2 \lambda_k$ and, since the terms of $\bar{z}$ are uncorrelated, we can use the Central Limit Theorem to approximate the distribution of this last expression with a normal distribution for a large value of $N$. The mean and variance of this distribution are given by:
\begin{align*}
    \mu &= (\sigma_\beta^{-2} - \sigma_\alpha^{-2})\sum_{k=1}^N \lambda_k
    \\
    &= (\sigma_\beta^{-2} - \sigma_\alpha^{-2}) \mathrm{tr}(\bar{\Sigma}_y) = N (\sigma_\beta^{-2} - \sigma_\alpha^{-2}) \sigma_y^2
    \\
    \sigma^2 &= (\sigma_\beta^2 - \sigma_\alpha^2)^2 \sum_{k=1}^N \lambda_k^2 = (\sigma_\beta^2 - \sigma_\alpha^2)\mathrm{tr}(\bar{\Sigma}_y \bar{\Sigma}_y)
\end{align*}
For the $\mu$, this is equal to $N \sigma_y^2$, so the expected value of the cumulative log-likelihood is not affected by the correlation between steps. For the variance, we use the fact that $\mathrm{tr}[\bar{\Sigma}_y \bar{\Sigma}_y] = \sum_{i=1}^N \sum_{j=1}^N (\bar{\Sigma}_y \bar{\Sigma}_y)_{i j}^2$ since $\bar{\Sigma}_y=\bar{\Sigma}_y^\top$ and get its asymptotic rate of increase when $N\rightarrow \infty$:
\begin{align*}
    &\frac{\sigma^2}{\left(\sigma_\alpha^{-2} - \sigma_\beta^{-2} \right)^2}
    =
    \mathrm{tr}[\bar{\Sigma}_y \bar{\Sigma}_y]
    =
    \sum_{i=1}^N \sum_{j=1}^N \left(\bar{\Sigma}_y\right)_{ij}^2 \nonumber
\end{align*}
As we are interested in asymptotic behavior, we compute how the last expression increases with $N$ as $N\rightarrow \infty$ by taking the limit:
\begin{align*}
    &\frac{1}{N} \sum_{i=1}^N \sum_{j=1}^N \left(\Sigma_y\right)_{ij}^2
    = \sigma_y^4 \left[ N + 2 \sum_{i=1}^N \lambda^i \left(N - i\right)\right]
    \\ &\stackrel{N\rightarrow \infty}{=} \sigma_y^2 N \left[1 + 2\sum_{i=1}^N \lambda^i\right]
    = \sigma_y^4 N\left[
        1 + 2 \lambda^2 \sum_{i=0}^N \lambda^{2i}
    \right] \nonumber
    \\ &= \sigma_y^4 N\left[1 + \frac{2 \lambda^2}{1-\lambda^2} \right]
    = N\sigma_y^4 \frac{1 + \lambda^2}{1-\lambda^2}
\end{align*}
The correlation between measurements, then, introduces a factor $(1+\lambda^2)/(1-\lambda^2)$ to the covariance, increasing it and therefore spreading the distribution of the cumulative log-likelihood ratio.

\bibliographystyle{IEEEtran}
\bibliography{./cdc26}

\end{document}